\documentclass[aps,amsfonts,pra,onecolumn,superscriptaddress,showpacs]{revtex4}
 
\usepackage{graphicx}
\usepackage{dcolumn}
\usepackage{bm}
\usepackage{natbib}
\begin{document}

\title{On the ubiquity of the L\'evy integral, its relationship with the generalised
Euler-Jacobi series, and their asymptotics beyond all orders.} 

\author{T.M. Garoni}
\email[]{t.garoni@physics.unimelb.edu.au}
\affiliation{School of Physics, University of Melbourne, Parkville,
Victoria 3010, Australia}
\author{N.E. Frankel}
\email[]{n.frankel@physics.unimelb.edu.au}
\affiliation{School of Physics, University of Melbourne, Parkville,
Victoria 3010, Australia}

\date{\today}
\newcommand{\ds}{\displaystyle}
\newcommand{\ts}{\textstyle}
\newcommand{\be}{\begin{equation}}
\newcommand{\ee}{\end{equation}}
\newcommand{\ba}{\begin{eqnarray}}
\newcommand{\ea}{\end{eqnarray}}
\newcommand{\bi}{\bibitem}
\newcommand{\intl}{\int\displaylimits}
\newcommand{\suml}{\sum\displaylimits}
\newcommand{\prodl}{\prod\limits}
\newcommand{\cupl}{\cup\displaylimits}
\newcommand{\p}{p_{\alpha}^d(r)}
\newcommand{\f}{F_{\alpha}(z)}
\newcommand{\pq}{p_{{p\over q}}^d(r)}
\newcommand{\levy}{{L\'evy }}
\newcommand{\polya}{{P\'olya }}
\newcommand{\xl}{<x^{2m}>_{\lambda}}
\newcommand{\modr}{\vert {\bf r}\vert}
\newcommand{\m}{m_{\alpha}^d(\r)}
\newcommand{\vs}{vs}
\newcommand{\LEVY}{L\'EVY }
\newcommand{\pone}{p_{\alpha}^1(r)}
\newcommand{\ponq}{{p\over q}}
\newcommand{\qonp}{{q\over p}}
\newcommand{\peven}{p_{2n\over q}^d(r)}
\newcommand{\g}{{p\over 2}}
\newcommand{\const}{{(2\pi)^{p-2-d\over 2}\over p^{\s +1}}}
\newcommand{\pn}{p_{2n}^d(r)}
\newcommand{\x}{\left({r^p\over p^p}\right)}
\newcommand{\D}{{-{1\over 2\pi r}}{\partial\over \partial r}}
\newcommand{\pp}{p_p^d(r)}
\newcommand{\bl}{\left(}
\newcommand{\br}{\right)}
\newcommand{\smallk}{k}
\newcommand{\phidp}{\phi^d_p}
\newcommand{\X}{{\bigwedge_{m=0}^{\infty}}}
\newcommand{\fpq}{F_{p/q}(z)}
\newcommand{\s}{S_{p/q}(a)}
\newcommand{\brezin}{Br\'ezin }
\newcommand{\hphi}{{\hat \phi}}
\newcommand{\hpsi}{{\hat \psi}}

\begin{abstract}
We present here an overview of the history, applications and important
properties of a function which we refer to as the \levy integral. For
certain values of its characteristic parameter the \levy integral
defines the symmetric \levy stable probability density function. As
we discuss however the \levy integral has applications to a number of
other fields besides probability, including random matrix theory,
number theory and asymptotics beyond all orders. We exhibit a direct relationship
between the \levy integral and a number theoretic series which we
refer to as the generalised Euler-Jacobi series. The complete
asymptotic expansions for all natural values of its parameter are
presented, and in particular it is pointed out that the intricate exponentially small series
become dominant for certain parameter values. 
\end{abstract}
\maketitle

 \begin{center}{\em Dedicated to the memory of Elliott Montroll}
 \end{center}

\section{Introduction}
The objective of this article is to pay homage to an
integral which has
occupied the minds of a host of great mathematicians including
Cauchy, L\'evy, Hardy, Littlewood, Ramanujan, and \polya and which
continues to arise in modern contexts such asymptotics beyond all
orders and  random matrix theory. This function, $\f$, which we refer
to as the  \levy integral is defined via
\be
\f=\int_{0}^{\infty}\exp(-t^{\alpha})\cos(z t)\, dt \qquad \alpha>0
\label{f}
.\ee
For $\alpha<1$ the integral (\ref{f}) is defined only for real
$z$.

The number of seemingly disparate topics in which $\f$ appears in a fundamental
way is remarkable. The reader familiar with the theory of probability
will immediately recognise the function $(1/\pi)\f$ as defining the symmetric \levy stable
probability density function when $\alpha\le 2$. For $\alpha =1,2$ the function
$(1/\pi)\f$ is identified respectively with the Cauchy and Gaussian probability
densities. However, $\f$ also has an intimate connection to Waring's
problem in number theory, is of great interest in
asymptotics beyond all orders, and appears in certain problems arising
in random matrix theory and wave phenomena.

The analytic properties of $\f$ are of long standing interest. 
Bernstein \cite{bernstein} in 1919 investigated the zeros of
$\f$, showing that it possesses  an infinite number of real zeros when
$\alpha =4,6,8,...$. He appears to have been the
first to realise that $\f$ might become negative for some values of
$z$ when $\alpha>2$. In such cases $(1/\pi)\f$ does not define a probability
density function. In 1923
\polya \cite{polyamathmess,polyaszego} extended the work of Bernstein by
systematically investigating the zeros of $\f$ and showed that when $\alpha =4,6,8,...$, $\f$ has an
infinite number of real zeros but no complex zeros, whereas if
$\alpha>1$ is not even integer then $\f$ has an infinite number of
complex zeros and a finite number, not less than $2[\alpha /2]$, of
real zeros. \levy \cite{levycr23} in 1923 proved 
the non-negativity of $\f$ for all real $z$ when $0<\alpha\le 2$, and was
responsible for the development of the theory of stable
distributions \cite{levybook}, and so we refer to the function $\f$ as the \levy
integral for all positive $\alpha$ regardless of whether it defines a
probability density function or not. It has been observed \cite{levy1,norm}
that for all positive rational $\alpha$ the function $\f$ can be expressed as a sum of generalised
hypergeometric functions, and in fact $\f$ is itself a specific type of
Fox H-function \cite{schneider}. There are however only a handful of specific
values of $\alpha$ for which an identification with more familiar
special functions is known \cite{levy1,zolotarev}.  

The present authors \cite{levyd} have recently constructed the complete asymptotic
expansions of $\f$ for all natural $\alpha$, and in particular have
demonstrated that there exist very intricate series of
exponentially small terms lying beyond all orders of the asymptotic
power series. Interestingly, when $\alpha$ is an even integer the
asymptotic power series vanishes and these exponentially small series
are liberated from their subdominance. This provides a powerful case
study for the newly emerging field of asymptotics beyond all orders\cite{dingle,norm},
the discussion of which we shall return to  after discussing how $\f$
arises in the fields of probability, random matrix theory,
diffraction integrals, and number theory.

\section{\LEVY stable probability density functions}
The theory of stable distributions is a branch of probability theory
of fundamental importance. At a time when it was fashionable to find
the most general conditions possible in order for the central limit
theorem to be satisfied, Paul \levy was interested in finding elegant
examples of when it is not. Indeed \levy's stable probability density
functions have divergent moments for $\alpha\ne 2$, and the definition
of stability is manifestly one of self similarity rather than universality.
\levy defines a probability density function $p(x)$ to be stable if it
satisfies the following property \cite{levycr23,levybook,barry}. 
Let $X_1$ and $X_2$ be independent random variables with the same
probability density function $p(x)$, then 
\be
c_1 X_1 + c_2 X_2 = cX
\ee
for all positive $c_1$ and $c_2$, where $X$ has the same probability
density function as $X_1$
and $X_2$. The constant $c$ is determined by the constants $c_1$ and $c_2$.
The symmetric solutions of this condition have characteristic functions
of the form $\exp(-{\vert { q}\vert}^{\alpha})$ \cite{barry}, which is the Fourier transform of $(1/\pi)\f$.

The function $\f$ was studied by Cauchy in the
early 1850s \cite{cauchy} in relation to the theory of errors, and it
is evident that he was aware that these functions satisfy \levy stability. Cauchy appears unaware however that $\f$ may become
negative for some values of $z$ when $\alpha>2$ and so fail the
non negativity requirement of a probability density function. 

The divergence of the moments of $\f$ for $\alpha$ not an even natural
can immediately be seen from the non-analyticity of $\exp(-{\vert
{ q}\vert}^{\alpha})$, since if the characteristic function of a
probability density function is analytic then it possesses a power
series whose coefficients are proportional to  the moments of that
probability density function. We can summarise\cite{levy1} the behaviour of the
moments of $(1/\pi)\f$ as a function of $\alpha$ as  
\be
<z^{2m}>=
\cases{
{(-1)^{m+j}(2m)!\over j!}, &$\alpha=2{m\over j}\in\{2,4,6,...\}$\cr
+\infty, &$\alpha<2m$ and $\alpha\in\cup_{k=0}^{\infty}(4k,4k+2)$\cr
-\infty, &$\alpha<2m$ and $\alpha\in\cup_{k=0}^{\infty}(4k+2,4k+4)$\cr
0, &otherwise.\cr}
\ee   
The stable probability density
functions have been experiencing a tsunami of popularity in a wide
variety of applications in the past decade, from physics to biology to
economics \cite{shlesinger}. The reason for their utility stemming
precisely from the divergence of the moments for $\alpha\ne 2$. 
This implies that  for $\alpha<2$ no characteristic
scale exists and random walks constructed using $(1/\pi)\f$ as the
step length probability density function display statistically self
similar trajectories, and as stressed by Mandelbrot the fractal  dimension of these
trajectories is $\alpha$\cite{mandelbrot,montroll1,montroll2}. Thus $\f$ plays a fundamental role in the
increasingly popular pursuit of modeling phenomena which display
structure on many scales. 

\section{Random matrix theory and the cusp diffraction catastrophe}
It has been noted \cite{levy1} that $F_4(y)$ is proportional to the
special case of the Pearcey integral $P(x,y)$ in which its first
argument is set to zero. Specifically, 
\be
P(0,y)= 2 e^{i\pi/8} F_4(y).
\ee

The Pearcey integral is the canonical form of the oscillatory integral
describing the cusp diffraction catastrophe \cite{paris,kaminski,wong} and is
defined by
\be
P(X,Y)= 
\int_{-\infty}^{\infty}\exp[i(u^4 +X u^2 + Y u)]\; du,
\ee
which can be expressed in an alternative more general form via
analytic continuation
\be
P(x,y)=
 2 e^{i\pi/8}
\int_{0}^{\infty}\exp(-t^4 -x t^2)\cos(y t)\; dt.
\ee 
This integral was investigated by Pearcey in his investigation of the
electromagnetic field near a cusp \cite{pearcey} and is  utilised in
many short wavelength problems such as wave propagation and optical
diffraction \cite{paris,kaminski,wong}. The function $P(x,y)$ was known before Pearcey
however, and was discussed by Lord Rayleigh in 1879 in connection with the
diffraction of light \cite{rayleigh}. Brillouin, as early as 1916, obtained an asymptotic expansion of
$P(x,y)$ using steepest descents \cite{brillouin}. As discussed in
\cite{paris}, $P(0,y)$ displays dominant exponentially small
oscillatory behaviour. As we shall discuss later, this behaviour is
shared by $\f$ for all even natural $\alpha$ larger than $2$.

\brezin and Hikami \cite{brezin} have recently demonstrated that
$F_4(z)$ appears in the solution of a certain problem in random matrix
theory. They consider an $N\times N$ Hamiltonian matrix, $H=H_0+V$,  which is
the sum of a given nonrandom Hermitian matrix, $H_0$, and a random Gaussian
Hermitian matrix $V$. The addition of $H_0$
breaks the unitary invariance of the probability measure that defines
the Gaussian unitary ensemble,
\be
P(H)\propto e^{-(N/2)TrV^2}\propto e^{-(N/2)Tr(H^2-2H_0H)}.
\ee
They investigate the eigenvalue density of $H$ when the  eigenvalues
of the nonrandom matrix, $H_0$, are chosen to be
$\pm a$, each value being $N/2$ times degenerate. In this case when
$a>1$ a gap appears in the eigenvalue spectrum of $H$ which closes as $a\to 1$.

An important quantity in random matrix theory is the $n$-point
correlation function, $R_n(\lambda_1,\lambda_2,...,\lambda_n)$, 
given by
\be
R_n(\lambda_1,\lambda_2,...,\lambda_n)=\left<\prod_{j=1}^{n}{1\over N}Tr\delta(\lambda_j-H)\right>,
\ee
where $\lambda_j$ is an eigenvalue of $H$.

The presence of nonzero $H_0$ implies that 
the usual orthogonal polynomial methods are no longer available for the
determination of the $n$-point correlation function. \brezin  and Hikami show however 
that the function $R_n(\lambda_1,\lambda_2,...,\lambda_n)$ is again expressible as the determinant
of an $n\times n$ matrix whose elements are given by a kernel
$K_N(\lambda_i, \lambda_j)$, just as in standard random matrix
theory.  The 
following contour integral expression for the kernel is arrived at

\ba
K_N(\lambda_1, \lambda_2) 
&=& 
(-1)^{(N/2)+1}
\int_{-\infty}^{\infty}{dt\over 2\pi} \oint_{\vert u\vert > a}{du\over
2\pi i}\left({t^2+a^2}\over u^2 - a^2\right)^{N/2}
\\ \nonumber 
&\times&
{1\over u-it}e^{-(N/2)u^2-(N/2)t^2 -Nit\lambda_1+Nu\lambda_2}.
\ea   

For the critical value of $a=1$ \brezin and Hikami then find that in
the scaling limit in which $N$ is large and $N^{3/4}\lambda$ is finite,
the kernel becomes
\be
{\hat K}(x,y)
=
{\hphi'(x)\hpsi'(y)-\hphi''(x)\hpsi(y)-\hphi(x)\hpsi''(y)\over x-y},
\ee  
where 
\be
\hphi(x)={\sqrt{2}\over \pi}F_4(\sqrt{2}x)
,\ee
and
\be
\psi(x)=
-{1\over \sqrt{\pi}}\sum_{n=0}^{\infty}{(-1)^n x^{4n+1} (2n)!\over n!(4n+1)!}
\ee
They then go on to show that this behaviour is universal
in the sense that ${\tilde K}(x,y)$ is independent of the eigenvalue
distribution of the non random Hermitian matrix $H_0$ provided that a gap
closes at the origin in the eigenvalue density of the total matrix $H$. 

\brezin and Hikami then discuss a situation in which one considers a
random Hermitian matrix coupled to a non random complex matrix,
${\tilde H}_0$,
rather than an Hermitian matrix. In this case the probability measure of the total matrix
becomes complex and so loses its probabilistic interpretation.
By appropriately selecting the eigenvalues of this non random matrix
\brezin and Hikami discuss a multicritical generalisation of the
closing gap phenomena discussed above, the kernel of which is
now expressible in an analogous manner in terms of $F_{2k}(z)$, with
each natural $k$ larger than $2$ corresponding to a particular choice of the eigenvalues of ${\tilde H}_0$
. As we mentioned in the introduction, and will
discuss in more detail presently, it is precisely this subclass of the
functions $\f$ that display dominant transcendental asymptotics.      

\section{The \LEVY integral, the generalised Euler-Jacobi series and
Waring's problem.}
Waring's conjecture \cite{hardywright} asserts that every natural number can be expressed
as the sum of $s$ integers all to the $k$th power. Let the function $g(k)$
denote the least value of $s$ such that the proposition is true for all natural numbers, and
$G(k)$ the least value of $s$ such that the proposition is true for
all but a finite number of natural numbers. Waring's problem naturally
decomposes into two parts, that of proving the existence of $g(k)$ for all
$k$, and that of the determination of the explicit structure of $g(k)$
and $G(k)$. Hilbert\cite{hilbert} proved the existence of $g(k)$ in 1909, and although
much progress took place last century \cite{hardywright} the
investigation of $g(k)$ and $G(k)$ continues to this day.

A fundamental arithmetic function that arises in the investigation of
Waring's problem \cite{hardywright,hardylittlewood} is $r_{k,s}(n)$ which for $k$
even is the number of representations of $n\in{\mathbb N}$ in the form
\be
n= a_1^k +a_2^k +...+ a_s^k \qquad a_i \in {\mathbb Z},
\ee
with representations which differ only in the order of the $a_i$
reckoned as distinct. The generating function of the arithmetic
function $r_{k,s}(n)$ which was investigated by Hardy and Littlewood \cite{hardylittlewood}
in relation to Waring's problem, takes the elegant form

\be
\sum_{n=0}^{\infty}
r_{k,s}(n) e^{-a n}
= 
[2 S_{k}(a)-1]^s 
\ee
where the function $S_{k}(a)$ is special case of a function which we refer to as the generalised Euler-Jacobi
series, defined by
\be
S_{p\over q}(a)= \sum_{n=0}^{\infty}e^{-a n^{p/q}}
\qquad p,q\in {\mathbb N}
\qquad \Re(a)>0.
\ee

The series $S_{p\over q}(a)$ was studied in detail in
\cite{norm}. Application of Euler's summation formula \cite{knopp,norm,levy1}
results in the following inversion, which is the generalisation
of the familiar transformation identity for the Jacobi theta function ${\cal\theta}_3(a)=2S_2(a)-1$.

\be
S_{p\over q}(a)=
{\Gamma\left(q/p +1\right)\over a^{q/ p}}
+{1\over 2}
+{2\over a^{q\over p}}
\sum_{n=1}^{\infty}F_{p/q}\left(2n\pi\over a^{q/p}\right).
\label{jacobilevy}
\ee

Hence we see that there exists a startlingly direct relationship between the number theoretic
Euler-Jacobi series $\s$, and the \levy integral $\fpq$.
For this reason, Hardy and Littlewood devoted an entire section of \cite{hardylittlewood}
to the investigation of $\f$, where it is noted that $F_{4}(z)$ can be
written as a sum of hypergeometric functions and that it displays dominant
exponential asymptotics, two observations which we show are indeed
true for all even natural $\alpha$. Without the modern theory of the generalised
hypergeometric functions at their disposal, as appears in \cite{luke}
for instance, an investigation along the hypergeometric route however
is dismissed by Hardy and Littlewood as
containing too many formal complications. In \cite{norm} the
hypergeometric representations are given for $\fpq$ and hence $\s$,
and the complete asymptotic expansions are investigated for various
cases of ${p\over q}$. Ramanujan  was interested in the series $\s$
and derived a small $a$ asymptotic expansion for it as discussed by
Berndt \cite{ramanujan}. This expansion consists only of the component
power series however which vanishes identically when $\alpha$ is an
even natural, all exponentially small terms having been
neglected. Berndt uses the Mellin transform technique rather than
utilising the inversion formula (\ref{jacobilevy}) and  hypergeometric representation of
$\f$ to recover Ramanujan's result. 

\section{The exponential asymptotics of the \LEVY integral}
The question of the asymptotics of $\f$
appears to have been first investigated thoroughly by Burwell
\cite{burwell}, who utilised the technique of steepest descents to
obtain the first terms in the dominant asymptotic expansion but was
not concerned with obtaining the exponentially small subdominant terms. 
By expressing $\f$ as a sum of generalised hypergeometric functions
the complete large $z$ asymptotic expansions of $\f$ were
calculated in \cite{norm} for $\alpha=3,4,5,6,7$, which then yielded
the small $a$ asymptotic expansions of $\s$. A more general
problem has recently been solved by the present authors. The $d$
dimensional isotropic \levy stable probability density function possesses 
the following integral representation, where $J_{(d/2) -1}(qr)$ is a Bessel function,

\ba
\p
& = &
{1\over (2\pi)^d}
\int_{{\mathbb R}^d}
d^dq\;
e^{-i{\bf r}\cdot {\bf q}}\exp(-{\vert {\bf q}\vert}^{\alpha})
\label{p formal}
\\
\p
& = &
{r\over (2\pi r)^{(d/2)}}
\intl_{0}^{\infty}dq\;
J_{(d/2) -1}(qr)
q^{(d/2)}
e^{-q^{\alpha}}
\label{p}.
\ea

When $d=1$ this clearly reduces to $(1/\pi)\f$. It is shown in
\cite{levyd} that $\p$ can be written in terms of generalised
hypergeometric functions for all positive rational $\alpha$. These
hypergeometric representations are then used to construct the  complete
asymptotic expansions for (\ref{p}) for all
natural $\alpha$. 
Utilising the useful notation, $f(\pm z)^{\pm}\equiv f(z) \pm
f(-z)$ for some function $f$, the \levy integral for rational $\alpha$
can be identified
with a sum of generalised hypergeometric functions as follows
\ba
F_{p/q}(z)
& = &
\pi q^{{q\over p} + {1\over 2}}\over p^{{3\over 2} }}{(2\pi)^{{(p-q-2)\over 2}} 
\sum_{l=0}^{p-1}(-1)^l\left({q^qz^p\over p^p}\right)^{2l\over p}
\nonumber \\
& \times &
{\prod_{h=1}^{(q-1)}\Gamma[{(2l+1)\over p} + {h\over q}] \over
 \prod_{h=1}^{(p-1)}\Gamma[{(2l+1)\over p} + {h\over p}]       }
{_qF_{p-1}}\left(
^{1,{2l+1\over p}+{1\over q},{2l+1\over p}+{2\over q},...,{2l+1\over
p}+{q-1\over q}}
_{{2l+2\over p},{2l+3\over p},...,{2l+p\over p}} 
\Bigg\arrowvert \pm i^p {q^qz^p\over p^p}\right)^+.
\label{fashyper}
\ea
where $p,q$ are natural numbers.
The complete asymptotic expansions of $\f$ and
$S_{\alpha}(a)$ for all natural $\alpha$ are a straight forward corollary of
the asymptotics for $\p$ given in \cite{levyd}, and we present the expansion for $\f$ below. In this way, this article can be
considered as an update of the work presented in \cite{norm}, in that
we now display the structure of the asymptotics for all
$\alpha\in{\mathbb Z}_{\ge 3}$, rather than discussing specific  cases.
We remark that the complete asymptotic expansion of $\p$ presented in \cite{levyd} is only extremely weakly
dependent on the dimension $d$, none of the overall structure changing
as one moves from $d=1$ to $d>1$.  Thus all of the interesting
asymptotic behaviour possessed by $\p$ is retained by $\f$ and in this
sense the asymptotic behaviour of $\p$ for natural $\alpha$ is
universal with regard to the spatial dimension $d$.  
\ba
\f
&\sim&
{\alpha\over z}
\sum_{m=1}^{\infty}
{(-)^{m+1}\over z^{\alpha m}}
{\Gamma(m\alpha)\over \Gamma(m)}
\sin\left({m\pi\alpha\over 2}\right)
\nonumber \\ 
&+&
\sqrt{{\pi\over 2(\alpha-1)}}
{2\over (\alpha z^{\alpha-2})^{1/2(\alpha-1)}}
\sum_{m=0}^{\infty}2^m N_m \left({\alpha\over z}\right)^{m\alpha/(\alpha-1)}
\nonumber \\ 
&\times&
\sum_{k=0}^{[\alpha/2]-[\alpha/4]-1}
\exp\left(-(\alpha-1)\sin\left[{(4k+1)\pi\over 2(\alpha-1)}\right]\left({z\over \alpha}\right)^{\alpha/(\alpha-1)}\right)
\nonumber \\
&\times&
\cos\left((\alpha-1)\cos\left[{(4k+1)\pi\over 2(\alpha-1)}\right]\left({z\over \alpha}\right)^{\alpha/(\alpha-1)}
-{(4k+\alpha)m\pi\over 2(\alpha-1)} + {(4k+2-\alpha)\pi\over 4(\alpha-1)}\right)
\nonumber \\
&-&
\sqrt{{\pi\over 2(\alpha-1)}}
{\delta_{(\alpha-4[\alpha/4]),2}\over (\alpha z^{\alpha-2})^{1/2(\alpha-1)}}
\sum_{m=0}^{\infty}2^m N_m \left({\alpha\over z}\right)^{m\alpha/(\alpha-1)}
(-)^m 
\nonumber \\ 
&\times&
\exp\left[-(\alpha-1)\left({z\over \alpha}\right)^{\alpha/(\alpha-1)}\right],
\label{fasymptotics}
\\
&-&
{\pi\over\alpha}<\arg(z)<{\pi\over\alpha}, 
\qquad\qquad \vert z\vert\to\infty,
\qquad\qquad \alpha \in {\mathbb Z}_{\ge 3}.
\nonumber 
\ea
Here $[x]$ refers to the integer part of the real number $x$, and
$\delta_{(\alpha-4[\alpha/4]),2}$ is the Kronecker delta. The $N_k$ are defined by the following recursion relation

\vbox{
\ba
N_k
&=&
\sum_{s=1}^{2(\alpha-1)}
\sum_{r=0}^{2(\alpha-1)-s}
\left({\alpha\over 2(\alpha-1)}(r+s-k)-(\alpha-1)+{\alpha\over 4(\alpha-1)}\right)_{\alpha-1}
\nonumber \\
&\times&
\left({\alpha\over 2(\alpha-1)}(r+s-k)-(\alpha-1)-{(\alpha-2)\over 4(\alpha-1)}\right)_{\alpha}
\nonumber \\
&\times&
{(-)^{s+r}[2(\alpha-1)]^{2\alpha-s-2}\over 
r![2(\alpha-1)-s-r]!\alpha^{2\alpha-1}k}N_{k-s},
\nonumber
\ea
\be
N_s =0 \qquad s<0,\qquad N_0=1.
\label{ns}
\ee
}
where $(x)_\alpha$ denotes the Pochhammer symbol.

We note that for the case of $\alpha=4$, (\ref{fasymptotics}) is in agreement with the corresponding result for
the Pearcey integral given in \cite{paris}. We remark also that
in the process of extending \polya's work on the zeros of $\f$, Senouf \cite{senouf}
has obtained the $k=0$ portion of the complete asymptotic expansion
(\ref{fasymptotics}) for even $\alpha$.

The small $a$ asymptotics of the series $S_{\alpha}(a)$ is now given simply by
utilising its relationship with the \levy integral (\ref{jacobilevy}).
Since the cases $\alpha=1,2$ are simply the Cauchy and Gaussian
probability density functions, up to a normalisation factor of $\pi$,
whose asymptotic behaviour is trivial to obtain, (\ref{fasymptotics}) gives us a
complete understanding of the asymptotic behaviour of $\f$ for all
natural $\alpha$. It is clear from the form of the power series
component  of the full asymptotic expansion, given by the first series
of (\ref{fasymptotics}), that when $\alpha$ is even every
term vanishes identically and so the exponentially small terms become
dominant. We note that when  $\alpha$
is an even integer the hypergeometric representation of $\f$
simplifies in such a way that only hypergeometric functions with no
numerator parameters appear\cite{levyd}, and it is known, see e.g. \cite{luke},
that such hypergeometric functions always display purely
transcendental asymptotics. Thus the function $\f$ displays qualitatively
different behaviour depending on whether the parameter $\alpha$ is
even or odd. Also, the presence of $[\alpha/4]$ in the summation
limits and the Kronecker delta show that the structure of the
asymptotics of $\f$ depends sensitively not only on the parity of
$\alpha$, but on the arithmetic residue of $\alpha$ modulo $4$. The
mechanism by which this dependence arises in the construction of
(\ref{fasymptotics}) is discussed in \cite{levyd}. An interesting property of (\ref{fasymptotics}) is
that the number of exponentially decaying oscillatory series increases
as $\alpha$ increases, so that the asymptotics becomes ever more
intricate as $\alpha$ becomes large. This phenomena was observed in
\cite{norm,levy1} for the specific cases $\alpha=3,4,5,6,7$, and predicted
to be true for general $\alpha$. Our result (\ref{fasymptotics}) now
confirms this. 

\section{Concluding Remarks}
The theory of stable distributions is a fundamental branch
of probability, and consequently the importance of the \levy integral for $\alpha$ less than two is 
widely recognised. What is equally as intriguing as its probabilistic
applications is the fact that the \levy integral also appears to be of
importance when $\alpha$ is larger than $2$ and it no longer defines a probability
density function. Its delicate asymptotics when
$\alpha$ is an even natural number is qualitatively different to anything
exhibited when $\alpha<2$. From examples such as its appearance in random matrix
theory, and its relationship with the Pearcey integral, one might well
expect that the future growth of interest in the \levy integral might
lie not in probability, in which it is well established, but in
problems in which its exponential asymptotics are utilised. In
particular it appears that the subclass of $\f$ where $\alpha$ is an
even natural number, which
exhibits dominant exponential asymptotics, will be of interest. 

\begin{acknowledgments}
 We would like to express our appreciation and indebtedness to Michael
 Shlesinger who introduced us to the delights of \levy flights.
 We would also like to thank Peter Forrester for numerous fruitful discussions on
 the theory of random matrices, and in particular for explaining to us
 the role of the \levy integral in random matrix theory.
\end{acknowledgments}


\end{document}